\documentclass[aip,rsi,amsmath,amssymb,reprint]{revtex4-1}
\usepackage{graphicx,dcolumn,bm}
\usepackage{comment}
\usepackage[utf8]{inputenc}
\usepackage[english]{babel}

\begin{document}

\preprint{AIP/123-QED}

\title{Sequences of the ranged amplitudes as a universal method for fast noninvasive characterization of SPAD dark counts}
\author{M.A. Smirnov}
\affiliation{Kazan Quantum Center, Kazan National Research Technical University n.a. A.N.Tupolev-KAI, 10 K. Marx, Kazan 420111, Russia}
\author{N.S. Perminov}
\affiliation{Kazan Quantum Center, Kazan National Research Technical University n.a. A.N.Tupolev-KAI, 10 K. Marx, Kazan 420111, Russia}
\affiliation{Zavoisky Physical-Technical Institute of the Russian Academy of Sciences, 10/7 Sibirsky Tract, Kazan 420029, Russia}
\author{R.R. Nigmatullin}
 \affiliation{Department of Radioelectronics and Information-Measuring Technique, Kazan National Research Technical University n.a. A.N.Tupolev-KAI, 10 K. Marx, Kazan 420111, Russia}
\author{A.A. Talipov}
 \affiliation{Kazan Quantum Center, Kazan National Research Technical University n.a. A.N.Tupolev-KAI, 10 K. Marx, Kazan 420111, Russia}
\author{S.A. Moiseev}
\altaffiliation{Author to whom correspondence should be addressed. Electronic mail: s.a.moiseev@kazanqc.org.}
\affiliation{Kazan Quantum Center, Kazan National Research Technical University n.a. A.N.Tupolev-KAI, 10 K. Marx, Kazan 420111, Russia}
\affiliation{Zavoisky Physical-Technical Institute of the Russian Academy of Sciences, 10/7 Sibirsky Tract, Kazan 420029, Russia}

\date{\today}

\begin{abstract}
Single-photon detectors based on avalanche photodiodes (SPAD) are key elements of many modern highly sensitive optical systems. One of the bottlenecks of such detectors is a afterpulsing effect which limits a detection rate and requires an optimal hold-off time. In this letter we propose a novel approach for statistical analysis of SPAD dark counts and we demonstrate its usefulness for the search of the experimental condition where the afterpulsing effect can be strongly eliminated. This approach exploits a sequence of ranked time intervals between the dark counts and does not contain a complex mathematical analysis of the experimental data. We show that the approach can be efficiently applied for a small number of the dark counts and it seems to be very beneficial for practical characterization of SPAD devices.
\end{abstract}

\maketitle

\textit{Introduction.}
Nowadays single-photon detectors based on avalanche photodiodes (SPAD) are actively used in quantum cryptography \cite{Zhang2015,Gisin2002,Hiskett2001}, laser ranging \cite{Wehr1999}, biological imaging \cite{Suhling2002}, optical quantum memory and entangled state measurement \cite{Ding2013,Mitchell2004}, single photon sources \cite{Eisaman2011,Chunnilall2014} and other sensitive devices in optical quantum information science \cite{Hadfield2009}.

It is possible to register an extremely low-level radiation due to the avalanche multiplication effect in avalanche photodiodes (APD).
One of the main problems of APDs is dark counts registered without incident photons. Unfortunately, it is difficult to distinguish the dark counts from the input single photon events which reduce the signal/noise ratio.
There are three main mechanisms of noise carrier generation, which contribute an appearance of dark counts: thermal generation, tunneling current and re-emission of previously trapped carriers \cite{Tan2001}.
The latter circumstance is the cause of the afterpulsing effect \cite{Cova1991,Cova1996}.
Below we focus to the properties of this effect

The afterpulsing can be suppressed by reducing the number of carriers flowing through the APD in avalanche events.
One of the methods for the afterpulsing suppression is to keep the bias voltage on the diode below the breakdown voltage for some time (hold-off time) after the each avalanche even.
In this case, premature avalanche events do not occur because the carriers released from the traps do not cause new avalanche events.
However, this method reduces the maximum operating rate of the SPADs.
Afterpulsing counts become more noticeable when the detector is cooled, since lower temperature slows down the yield process of the trapped carriers \cite{Jensen2006,Meng2016}.

There are several methods for investigation of dark counts: time interval analysis \cite{Jensen2006}, the double gate method \cite{Zhang2009}, temporal distribution (background decay) \cite{Dalla_Mora2012,Para2015} and many other techniques for various experimental installations \cite{Gong2014,Liu2016,Wiechers2016,Xu2016} (see also review \cite{Zhang2015}).
The time interval analysis is widely used.
Its essence lies in the construction of a histogram for the time intervals between samples of a single-photon detector.
It is well-known that the light sources or the dark events having a Poisson distribution produce the histograms with exponential distribution of the time intervals.
Any deviation from the exponential shape indicates to the correlation between the detection events and to the imperfection of the detector behavior, respectively \cite{Becker2005}.
In practice, the afterpulsing counts are revealed on the histogram as a sharp peak in the time interval from zero to several microseconds for InGaAs/InP APD.
However, the construction of histograms providing a sufficiently complete information about SPAD requires too large volume of experimental data \cite{Humer2015} (about $10^6$ dark counts).
In addition, the histograms depend on partition of temporal interval, thus losing some information corresponding to the values inside each interval.

In this letter, we propose a new and more universal method for statistical analysis of dark counts, which is based on using a sequence of ranged amplitudes (SRA) \cite{Nigmatullin2003,Nigmatullin2010,Baleanu2010} of the time intervals detected between the detector dark counts.
For demonstration we have tested a widely utilized commercial InGaAs/InP SPAD.
This analysis allows us to find the operating regimes of the detector with suppressed afterpulsing effect that corresponds to the Poisson statistic of dark counts.
We discuss advantages of the proposed method in comparison with histogram method in terms of relative simplicity and noninvasiveness (excluding losses of any information) of data processing and the possibility of robust operation with a quite small set of experimental data (on the order magnitude $\sim10^3$).

\textit{Theoretical background.}
We study a free operation of SPAD without detection of external signal fields. Herein, time tags corresponding to the single dark counts are fixed. The time intervals between the dark counts of SPAD can be labeled as a sequence $\{X_i\}$, where the index "$i$" denotes the number in the sequence. Below we study the statistical properties of the dark counts by using a SRA approach. SRA data is constructed as follows from the initial data \cite{Nigmatullin2010}: each successive element $x_n$ in the SRA data $\{x_n\}$ is less than the previous one, i.e. $x_n\geq x_{n-1}$ (duplicate values are arranged in a series one after the other), where the first element $x_1$, respectively, has a maximum value of the time delay, and last element $x_N$ is the minimum value.
We assume that an approximate analytical expression that describes the resulting SRA curve can be found using its relation with the statistical distribution function $F(x,x_n)$.
The distribution function $F(x,x_n)$ describes the probability $P(x\leq x_n)$ that the value $x$ is less than (or equal) to $x_n$. Assuming no duplicate values in the sequence $\{x_n\}$, we can write the following expression:
\begin{align}\label{distr_func}
F(x,x_n)=P(x\leq x_n)=\underset{N\rightarrow\infty}{\operatorname{Lim}}\frac{N+1-n(x_n)}{N},
\end{align}
where $N$ is the total number of points in the sample (in our case $N=10^3$).

By taking into account that many light sources as well as the dark noise counts corresponding to the Poisson processes \cite{Drake1967}, the probability density for time intervals being between the counts will be approximately described by the damped exponent (see also theoretical review \cite{Wiechers2016,Horoshko2016}):
\begin{align}\label{hist}
\frac{dF}{dx}=\rho(x_n)=\lambda\operatorname{exp}(-\lambda x_n),
\end{align}
where $\lambda$ is the average frequency of avalanche events, and $x_n$ are the time intervals between counts. From the equations (\ref{distr_func}) and (\ref{hist}) we finally obtain for finite $N\gg1$:
\begin{align}\label{SRA}
x_n=\frac{1}{\lambda}\operatorname{ln}\left(\frac{N}{n-1}\right).
\end{align}
We use equation (\ref{SRA}) as a fitting function for SRA data for finite $N$, which has only a single fitting parameter $\lambda$. If $n=1$ the equation (\ref{SRA}) becomes infinity.
This point we exclude from consideration, i.e. will use in the future $n=2,3,...,N$.
For non-Poisson processes, the expression (\ref{SRA}) will not fit the SRA data but another fitting function can be found from equation (\ref{distr_func}) by using a corresponding distribution function $F(x,x_n)$.
We note that average time interval is linked with average frequency as $\langle x\rangle=1/\lambda$.
It leads to following relation $x_n/\langle x\rangle=\operatorname{ln}(N/(n-1))$.
Below, by using SRA approach and the obtained here universal relation for the Poisson process, we study statistical properties of SPAD dark counts and show that such a characterization can be highly efficient when the number of measured data is smaller in a three order of magnitudes in comparison with the well-known histogram method.
In our work, we used the ranking in descending order, and as the initial data we took a sequence of $10^3$ values of time intervals between dark samples.

\begin{figure}[htb]
\includegraphics[width=0.49\textwidth]{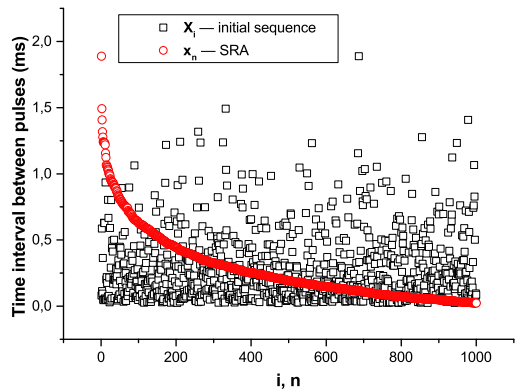}
\caption{An example of a sequence of time interval values $X_i$ (black empty squares) between dark samples and a sequence of ranged values of time intervals $x_n$ (red empty cycles).
The ranking was carried out in descending order.
Sequences contain $10^3$ values of time intervals.}
\label{fig_SRA}
\end{figure}

\textit{Data processing of SPAD dark counts.}
The data on dark count time intervals have been obtained from the commercial infrared SPAD based on InGaAs/InP APD (ID210, ID QUANTIQUE).
The data have been fixed in commercial time-to-digital converter (ID801, ID QUANTIQUE). Herein, we used the following SPAD parameters: mode -- internal gating, gate frequency -- 20MHz, effective gate width -- 3ns, efficiency -- 25\%.
In the course of the measurements, only one hold-off time tho was varied.
Then, the time intervals $X_i$ between the neighboring time-tags were calculated. 

\begin{figure}[htb]
\includegraphics[width=0.49\textwidth]{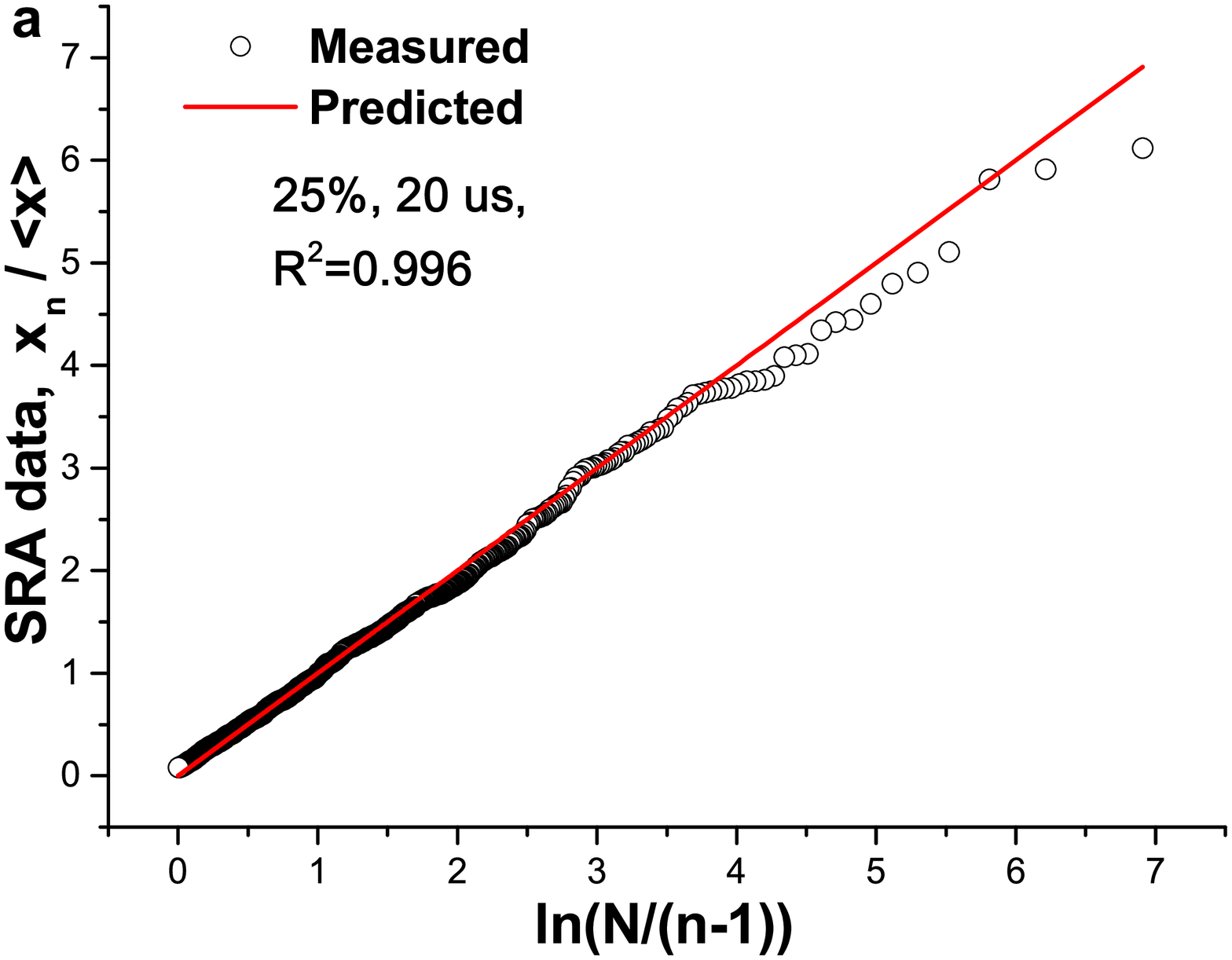}
\includegraphics[width=0.49\textwidth]{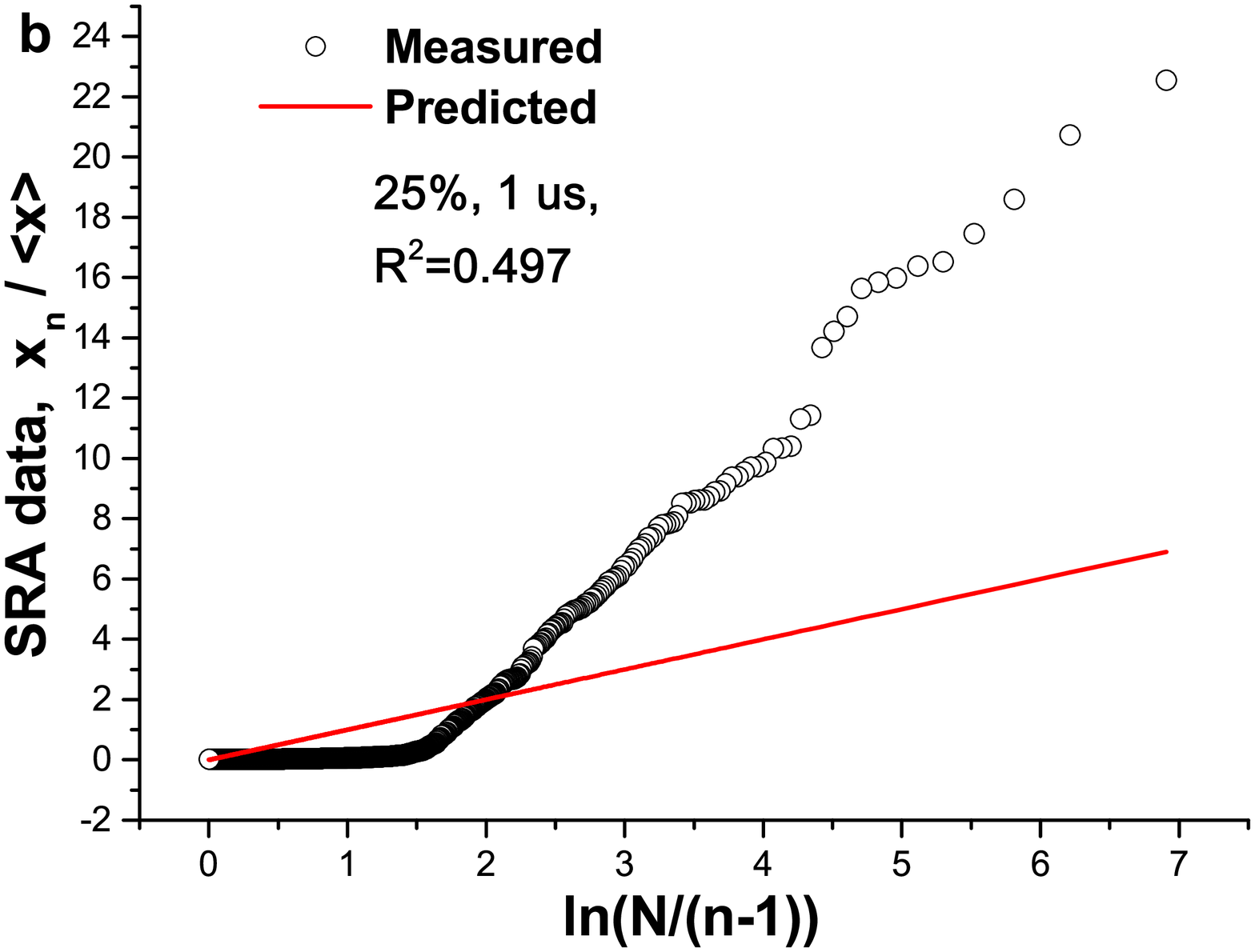}
\caption{Comparison of experimental SRA data obtained with a time $t_{ho}=20\mu$s (\textbf{a}) and $t_{ho}=1\mu$s (\textbf{b}) with the prediction data.
$N$ is the total number of points in the sample and equal $10^3$. 
To good visibility, both curves were plotted as function on $\operatorname{ln}(N/(n-1))$.}
\label{log_SRA}
\end{figure}

We chose a sequence of $10^3$ $X_i$ values.
The initial sequence of time intervals $X_i$ was further analyzed based on the described SRA data $\{x_n\}$ constructed for $\{X_i\}$.
Figure \ref{fig_SRA} shows a typical example of the $\{X_i\}$ and $\{x_n\}$ data.
One can see that SRA data form a smooth curve even for a small number of dark counts.
Herein, it was found that the shape of the curves depends on the hold-off time $t_{ho}$ of SPAD which was varied for different series of experiment. 

Figure \ref{log_SRA} compare experimental SRA data with theoretical equation (\ref{SRA}).
In figure \ref{log_SRA}\textbf{a}, the experimental data were obtained with a time $t_{ho}=20\mu$s.
This time is significantly longer than the time intervals corresponding the afterpulsing counts.
We see a quite good agreement between experimental data and theoretical curve.
However, an afterpulsing contribution is increased when the time interval $t_{ho}$ is reduced.
In this case, the experimental SRA data are deviated from the equation (\ref{SRA}), since we assumed a single Poisson process there.
Figure \ref{log_SRA}\textbf{b} demonstrates this behavior which has been obtained for the quite short time $t_{ho}=1\mu$s.
Additional measurements have shown that using independently obtained $10^3$ experimental counts gives a reproducible SRA curves indicating a high accuracy of this method for the such small number of counts.
The relative error in SRA data does not exceed 1\%.

\begin{figure}[htb]
\includegraphics[width=0.49\textwidth]{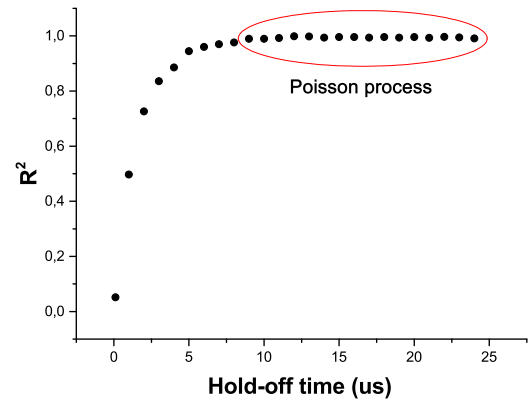}
\caption{Determination coefficient $R^2$ calculated for experimental SRA data.}
\label{det_coef}
\end{figure}

To estimate a concurrence between the experimental data and the theoretical curve given by equation (\ref{SRA}), we used the determination coefficient \cite{Rawlings1988} $R^2=1-\sum_i(y_i-f_i)^2/\sum_i(y_i-\langle y\rangle)^2$, where $y_i$ are experimental normalized values $x_n$, $f_i=\operatorname{ln}(N/(n-1))$ are the predicted values, $\langle y\rangle$ is a mean value of $y_i$.
The relation $R^2\cong1$ indicates to a high concurrence of experimental data with the theoretical one, i.e. with Poisson process.
By using this property of $R^2$, we have found a domain of the hold-off times $t_{ho}$ where the ideal Poisson process of dark counts occurs.
The dependence of the calculated $R^2$ on the time $t_{ho}$ is plotted in figure \ref{det_coef}. It is seen that $R^2$ is close to unity where the hold-off time is greater than 9$\mu$s. Whereas, at hold-off time values less than 9$\mu$s, the determination coefficient $R^2$ decreases quickly to zero.
These observations indicate a high sensitivity and productivity of the SRA approach for characterization Poisson process in dark counts of SPADs for small number of experimental data.

\textit{Conclusion.}
In the presented letter, we have proposed a new method for non-invasive statistical analysis of dark counts in SPADs, which is based on the SRA for time intervals between the dark counts.
The non-invasiveness of the SRA method allows to stay within the discrete statistics and without using artificial parameters.
In particular SPA eliminates  the partitioning of time intervals which is inherent to the method of histograms.
We have demonstrated the extreme precision and robustness of the 
SRA approach for a relatively small number ($10^3$) of dark counts that indicates to its advantages and simplicity in practical characterization of  SPADs.

\textit{Acknowledgments.}
The authors thank K.I. Gerasimov for fruitful discussion.
This research has been supported by the Government of Russian Federation, project no. 14.Z50.31.0040, Feb. 17, 2017.

\nocite{*}
\bibliographystyle{apsrev4-1}
\bibliography{SRADC}

\end{document}